\documentclass[a4paper, 11pt]{article}
\usepackage[utf8]{inputenc}
\usepackage[left=2cm, right=2cm, top=2.cm, bottom=2.cm]{geometry}
\pdfoutput=1

\usepackage{pifont}
\usepackage{slashed}
\usepackage{bm}
\usepackage{latexsym,amssymb,amsmath,float,url,mathrsfs,physics}
\usepackage{latexsym}
\usepackage{graphicx}
\usepackage{epstopdf}
\usepackage{amsmath}
\usepackage{comment}
\usepackage{subfigure}
\usepackage{pifont}
\usepackage{blindtext}
\usepackage{adjustbox}
\usepackage{multirow}
\usepackage{tabularx}
\usepackage[table,dvipsnames]{xcolor}
\usepackage{orcidlink}
\usepackage{tikz}
\usetikzlibrary{tikzmark}
\usepackage{fontawesome}

\usepackage{blkarray}
\usepackage{graphicx}
\usepackage{amsfonts}
\usepackage{soul}
\usepackage{amssymb}
\usepackage{cancel}
\usepackage{tcolorbox}
\usepackage{tensor}
\usepackage{textcomp}

\usepackage{graphics,appendix,afterpage,makecell}

\usepackage[
  backend=biber,
  style=numeric-comp,
  giveninits=true,
  doi=false,
  sorting=none,
  backref,backrefstyle=none,
  maxnames=7,
  minnames=6
]{biblatex}
\AtBeginBibliography{\small}  
\DeclareFieldFormat[article]{title}{\textit{#1}} 
\DeclareFieldFormat[incollection]{title}{\textit{#1}} 
\DeclareFieldFormat[conference]{title}{\textit{#1}} 
\DeclareFieldFormat[inproceedings]{title}{\textit{#1}} 
\DefineBibliographyStrings{english}{ 
  backrefpage = {p\adddot\addspace}, 
  backrefpages = {pp\adddot\addspace} } 
 
\renewbibmacro{in:}{%
  \ifentrytype{incollection}{%
   \printtext{\bibstring{in}\intitlepunct}%
   }{\addcomma\space }%
   }
\renewbibmacro*{doi+eprint+url}{
  \setunit{\addcomma\addspace}\newblock 
  \iftoggle{bbx:doi}
    {\printfield{doi}}
    {\printfield{doi}}%
  \setunit{\addcomma\addspace}\newblock 
  \iftoggle{bbx:eprint}
    {\usebibmacro{eprint}}
    {}%
  \setunit{\addcomma\addspace}\newblock
  \iftoggle{bbx:url}
    {\usebibmacro{url+urldate}}
    {}}
\DeclareFieldFormat{doi}{%
  \ifhyperref
    {\href{https://doi.org/#1}{\textsc{doi}}}
    {doi}}
\makeatletter
\DeclareFieldFormat{eprint:arxiv}{%
  arXiv\addcolon\space
  \ifhyperref
    {\href{http://arxiv.org/\abx@arxivpath/#1}{%
       #1
       \iffieldundef{eprintclass}
     {}
     {\addspace\mkbibbrackets{\thefield{eprintclass}}}}}
    {#1
     \iffieldundef{eprintclass}
       {}
       {\addspace\mkbibbrackets{\thefield{eprintclass}}}}}
\makeatother
\renewbibmacro*{publisher+location+date}{%
  \printlist{publisher}%
  \iflistundef{location}
    {\setunit*{\addcomma\space}}
    {\setunit*{\space(}}%
  \printlist{location}%
  \setunit*{)\addcomma\space}%
  \usebibmacro{date}%
  \newunit}
\DeclareSourcemap{
 \maps[datatype=bibtex,overwrite=true]{
  \map{
    \step[fieldsource=Collaboration, final=true]
    \step[fieldset=usera, origfieldval, final=true]
  }}}
\renewbibmacro*{author}{%
  \iffieldundef{usera}{%
  \printnames{author}%
  }{%
  \textbf{\printfield{usera}} collaboration, \printnames{author}%
  }}

\usepackage{url}
\def\lsim{\mathrel{\rlap{\lower4pt\hbox{\hskip0.5pt$\sim$}}
 \raise1pt\hbox{$<$}}}         
\def\gsim{\mathrel{\rlap{\lower4pt\hbox{\hskip0.5pt$\sim$}}
 \raise1pt\hbox{$>$}}}         

\renewcommand{\dd}{\mathrm{d}}
\newcommand{\etat}{\widetilde{\eta}}
\newcommand{\vx}{\mathbf{x}}
\newcommand{\vk}{\mathbf{k}}

\newcommand{\calH}{\mathcal{H}}

\newcommand{\Pz}{\mathcal P_\zeta}

\newcommand{\OGW}{\Omega_\text{GW}}
\newcommand{\rhoGW}{\rho_\text{GW}}

\newcommand{\Rs}{R_s}

\usepackage[normalem]{ulem}

 \def\be   {\begin{equation}}   \def\ee   {\end{equation}}
 \def\ba   {\begin{array}}      \def\ea   {\end{array}}
 \def\bea  {\begin{equation}\begin{aligned}}   \def\eea  {\end{aligned}\end{equation}}
 \def\bean {\begin{eqnarray*}}  \def\eean {\end{eqnarray*}}

\definecolor{TardisBlueAcceso}{RGB}{0, 80, 200}

\definecolor{oucrimsonred}{rgb}{0.6, 0.0, 0.0}
\definecolor{persianblue}{rgb}{0.11, 0.22, 0.73}
\definecolor{forestgreen}{rgb}{0.13,0.35,0.13}
\definecolor{lightgray}{rgb}{0.83, 0.83, 0.83}
\definecolor{oucrimsonred}{rgb}{0.6, 0.0, 0.0}

\definecolor{verdes}{rgb}{0.1, 0.5, 0.1}%
\definecolor{cornellred}{rgb}{0.7, 0.11, 0.11}

\definecolor{VioletRed4}{rgb}{0.55, 0.13, .32}
\definecolor{rossocorsa}{rgb}{0.83, 0.0, 0.0}

\usepackage{hyperref}
\hypersetup{colorlinks, citecolor=ForestGreen, linkcolor=BrickRed, urlcolor=NavyBlue}
\usepackage[capitalise]{cleveref}

\begin{document}

\title{\bf Understanding the Nature \\of Scalar-Induced Gravitational Waves}

\author{A.J. Iovino$^{1,2}$\orcidlink{0000-0002-8531-5962},
G.Perna$^{3,4,5,2}$\orcidlink{0000-0002-7364-1904}, D. Perrone$^2$\orcidlink{0000-0003-4430-4914}, D.Racco$^{6,7}$\orcidlink{0000-0002-0859-8751}, A. Riotto$^2$\orcidlink{0000-0001-6948-0856}}

\date{} 
\maketitle

\vspace{-14em}
\begin{flushright}%
ZU-TH 63/25%
\end{flushright}%
\vspace{4em}
\vspace{1.5cm}
\begin{center}
{\small
$^1$~New York University, Abu Dhabi, PO Box 129188 Saadiyat Island, Abu Dhabi, UAE\\
$^2$~Department of Theoretical Physics and Gravitational Wave Science Center, \\24 quai E. Ansermet, CH-1211 Geneva 4, Switzerland\\
$^3$~Dipartimento di Fisica e Astronomia ``Galileo Galilei'', Universit\`a degli Studi di Padova, \\Via Marzolo 8, I-35131, Padova, Italy\\
$^4$~INFN, Sezione di Padova, Via Marzolo 8, I-35131, Padova, Italy\\
$^5$~Institute for Theoretical Physics, Leibniz University Hannover, \\Appelstraße 2, 30167 Hannover, Germany\\
$^6$~Physik-Institut, Universit\"at Z\"urich, Winterthurerstrasse 190, 8057 Z\"urich, Switzerland\\
$^7$~Institut f\"ur Theoretische Physik, ETH Z\"urich,Wolfgang-Pauli-Str.\ 27, 8093 Z\"urich, Switzerland
}
\end{center}
\begin{abstract} 
\noindent
We offer a   physical interpretation of the origin of the scalar-induced gravitational wave background, showing that it is  mainly produced around the peaks of the scalar perturbations. We also provide a compact expression to estimate the amount of scalar-induced gravitational waves generated by  peaks.

\end{abstract}
\section{Introduction}
A central objective of contemporary and forthcoming gravitational wave (GW) observatories is the search for scalar-induced gravitational waves (SIGWs). This stochastic background of gravitational radiation, which arises as a generic prediction of General Relativity, is generated by scalar perturbations that undergo amplification relative to the primordial seeds responsible for the formation of large-scale structure~\cite{Tomita:1975kj, Matarrese:1993zf, Acquaviva:2002ud, Mollerach:2003nq, Carbone:2004iv, Ananda:2006af, Baumann:2007zm} (for a recent review, see Ref. \cite{Domenech:2021ztg}). For the resulting signal to be observationally significant, these scalar fluctuations must be enhanced well beyond the amplitude inferred from cosmic microwave background measurements. A variety of early-universe scenarios naturally predict such an amplification of the curvature perturbation $\zeta$ on small scales, thereby giving rise to SIGWs with distinctive spectral characteristics. Prominent examples include single-field inflationary models featuring ultra-slow-roll phases~\cite{Ivanov:1994pa,Leach:2000ea,Bugaev:2008gw,Alabidi:2009bk,Drees:2011hb,Drees:2011yz,Alabidi:2012ex,
Ballesteros:2017fsr,Di:2017ndc,Germani:2017bcs,Cicoli:2018asa,Ozsoy:2018flq,Bhaumik:2019tvl,Ballesteros:2020qam,Karam:2022nym,Franciolini:2022pav,Balaji:2022rsy,Iovino:2024sgs,Allegrini:2024ooy},
multifield inflationary dynamics~\cite{Garcia-Bellido:1996mdl,Bugaev:2011wy,Kawasaki:2015ppx,Clesse:2015wea,Braglia:2020eai,Palma:2020ejf,Braglia:2022phb,Balaji:2022dbi},
and curvaton mechanisms~\cite{Kawasaki:2012wr,
Ando:2017veq,Ando:2018nge,Ferrante:2022mui,Ferrante:2023bgz,Kawasaki:2021ycf,Chen:2019zza,Liu:2020zzv,Inomata:2020xad,Pi:2021dft, Gow:2023zzp,Inomata:2023drn}.
The detection and subsequent characterization of SIGWs would constitute a powerful probe of the primordial universe, offering direct insight into the physical processes operating during its earliest epochs. In addition, these signals are intimately connected to the formation of primordial black holes (for a comprehensive review, see Ref.~\cite{LISACosmologyWorkingGroup:2023njw}).

 It is known that GWs may be produced only when a nonvanishing quadrupole is present in their  source. However, in the standard second-order calculation its presence is obscured by technicalities and one may  lose track of the basic physical origin of the GWs. The ultimate reason is that one expands (an assumption which is rarely outlined) around a ground state for which the average of the curvature perturbation $\zeta({\bf x},\eta)$ vanishes. However, this is not true in general. There are regions where the vacuum expectation value of the curvature perturbation is not vanishing. Those are the regions where there are peaks (maxima or minima) of the curvature perturbations, whose statistics has been beautifully described in Ref. \cite{Bardeen:1985tr}. Furthermore, peaks are generically aspherical, which allows an intuitive understanding of where the quadrupole arises from. Another way to understand why peaks should be relevant is the fact that the source of the SIGWs is proportional to $\zeta\partial_i\partial_j\zeta$, and therefore one might expect that the largest contributions originate from points where the perturbation and its second derivatives are  the largest, i.e. around peaks.

The goal of this paper is to understand the origin of the SIGW background and to estimate how much of it is due to   the peaks of the curvature perturbations. Besides offering a nice physical interpretation, we will be able to provide a simple and compact expression for the size of the  SIGW background.

The paper is organised as follows. In section \ref{Sec::SIGW_Summary} we briefly summarize the standard calculation of the SIGW background. In section \ref{Sec::Peaks} we report some basic relations about peak theory, while in section \ref{Sec::SIGW_Nature} we perform our estimate of the contribution of the SIGW from the fluctuations around the peaks.  Section \ref{Sec::Conclusions} contains our conclusions.

\section{Scalar-Induced Gravitational Waves: a brief summary}
\label{Sec::SIGW_Summary}
In this section we briefly summarize the physics of the SIGW background and fix some notations. The perturbed metric in the conformal Newtonian gauge is expressed as
\begin{equation}
\mathrm d s^2 = -a^2(1+2\Phi)\mathrm d\eta^2 + a^2\left[(1-2\Psi)\delta_{ij} + \frac{1}{2}h_{ij}\right] \mathrm dx^i \mathrm dx^j ,
\end{equation}
where $\Phi$ and $\Psi$ denote the Bardeen potentials, while $h_{ij}$ captures tensor fluctuations, constrained to be transverse and traceless: $\partial_i h_{ij} = h_{ii} = 0$. When anisotropic stress can be neglected, one has $\Phi = \Psi$. The tensor perturbation $h_{ij}$ can be decomposed over a polarisation basis $\{e^{(+)}_{ij}, e^{(\times)}_{ij}\}$ as
\begin{equation}
h_{ij}(\eta, \vx) = \int \frac{\mathrm d^3k}{(2\pi)^3} \left[h_{\vk}^{(+)}(\eta)e_{ij}^{(+)}(\vk) + h_{\vk}^{(\times)}(\eta)e_{ij}^{(\times)}(\vk)\right] e^{i\vk \cdot \vx}.
\end{equation}
The polarisation tensors are defined through the vectors $e_i(\vk)$ and $\bar e_i(\vk)$, both orthonormal to $\vk$, as
\begin{align}
e_{ij}^{(+)}(\vk) &= \frac{1}{\sqrt 2} \left[e_i(\vk)e_j(\vk) - \bar e_i(\vk)\bar e_j(\vk)\right], \\
e_{ij}^{(\times)}(\vk) &= \frac{1}{\sqrt 2} \left[e_i(\vk)\bar e_j(\vk) + \bar e_i(\vk) e_j(\vk)\right],
\end{align}
and satisfy the normalisation and orthogonality conditions $e_{ij}^{(s)} e_{ij}^{(s')} = \delta^{ss'}$ for $s,s' = (+), (\times)$.
The evolution equation for GWs is obtained by isolating the tensor contribution in Einstein's equations, expanded to second order in perturbations
\begin{equation}
h_{ij}'' + 2\mathcal H h_{ij}' - \nabla^2 h_{ij} = -4 \mathcal T_{ij}{}^{lm} \mathcal S_{lm},
\label{eq: eom GW1}
\end{equation}
where primes denote derivatives with respect to conformal time, $\mathcal H = a'/a$ is the conformal Hubble rate, and $\mathcal S_{lm}$ represents a given  source term. The operator $\mathcal T_{ij}{}^{lm}$ projects out the transverse-traceless component of the source.

The  projector in Fourier space becomes
\begin{equation}
{\mathcal T}_{ij}{}^{lm}(\vk) = e_{ij}^{(+)}(\vk) e^{(+)lm}(\vk) + e_{ij}^{(\times)}(\vk) e^{(\times)lm}(\vk).
\end{equation}
We adopt the Fourier transform convention:
\begin{equation}
\mathcal S_{lm}(\eta, \vx) = \int \frac{\mathrm d^3k}{(2\pi)^3} {\cal S}_{lm}(\eta, \vk) e^{i\vk \cdot \vx},
\end{equation}
so that the equation of motion~\eqref{eq: eom GW1} translates, for each polarisation $s = +, \times$, to
\begin{equation}
{h_{\vk}^s}''(\eta) + 2 \mathcal H \, {h_{\vk}^s}'(\eta) + k^2 h_{\vk}^s(\eta) = {\cal S}^s(\eta, \vk),
\label{eq: eom GW2}
\end{equation}
with ${\cal S}^s(\eta, \vk) \equiv -4 e^{s,lm}(\vk){\cal S}_{lm}(\eta, \vk)$  the Fourier transform of the  source. This inhomogeneous differential equation can be solved via the Green's function method
\begin{equation}
h_{\vk}^s(\eta) = \frac{1}{a(\eta)} \int^\eta \dd \etat \, g_{\vk}(\eta, \etat) \, a(\etat) \, {\cal S}^s(\etat, \vk),
\end{equation}
where the retarded Green function in a radiation-dominated background is given by:
\begin{equation}
g_{\vk}(\eta, \etat) = \frac{\sin[k(\eta - \etat)]}{k} \, \theta(\eta - \etat),
\label{eq: Green function}
\end{equation}
with $\theta$ denoting the Heaviside function.
The source ${\cal S}_{ij}$ driving the tensor perturbations in Eq.~\eqref{eq: eom GW1} arises from second-order contributions of the scalar potential $\Psi$. Its explicit expression in real space reads
\begin{equation}
\mathcal S_{ij} = 2\Psi \partial_i\partial_j \Psi + \frac{4}{3(1+w)} \left(\frac{\Psi'}{\mathcal H} + \Psi\right) \partial_i\partial_j\left(\frac{\Psi'}{\mathcal H} + \Psi\right),
\label{eq: source x space}
\end{equation}
where $w$ is the background equation of state. Since the generation of GWs is most efficient when relevant modes cross the Hubble radius, a process that, in our case of interest, takes place well inside the radiation-dominated (RD) era, we take $w = 1/3$. In Fourier space, the scalar potential $\Psi(\eta, \vk)$ is related to the comoving curvature perturbation $\zeta$ via
\begin{equation}
\Psi(\vk,\eta) = \frac{2}{3} T(k,\eta) \zeta(\vk),
\label{eq: Psi to zeta}
\end{equation}
where the transfer function during RD is given by
\begin{equation}
T(k,\eta)  =\frac{9}{k^2\eta^2}\left[\frac{\sin(k\eta/\sqrt{3})}{k\eta/\sqrt{3}} - \cos(k\eta/\sqrt{3})\right].
\label{eq: transfer}
\end{equation}
\noindent
Finally, the density parameter of GWs per logarithmic interval of $k$,  in terms of the GW power spectrum ${\cal P}_h(k)$ and the critical energy density reads, at the time of production and for wavenumbers well inside the Hubble radius,
\begin{equation}
\OGW(k)|_{\rm e} = \frac{\rhoGW(\eta,k)}{\rho_\text{cr}(\eta)}=\frac{1}{24} \left(\frac{k}{\calH(\eta)}\right)^2 {\cal P}_h(k,\eta).
\end{equation}
As a final comment, let us remark that in the standard calculation \cite{Espinosa:2018eve,Kohri:2018awv} the curvature perturbation represents  the fluctuation around a vanishing average,
\be
\langle\zeta({\bf x})\rangle=0.
\ee
Nevertheless, there are regions in space where the classical value of the curvature perturbation is nonvanishing; there are indeed peaks of the curvature perturbation which can be characterized by an average profile such that (a more detailed discussion about the definition of the peaks is provided in the next section)
\be
\langle\zeta({\bf x})\rangle=\zeta_{\rm pk}({\bf x}).
\ee 
The question we wish to answer in the following is how much of the SIGW background is contributed by the fluctuations around such peaks.

\section{Characterizing the peaks of the curvature perturbation}
\label{Sec::Peaks}
In this section we summarize some basic results we borrow from the monumental reference \cite{Bardeen:1985tr}. The expert reader may skip this section.
For simplicity we assume that our initial curvature perturbation $\zeta$ is a random Gaussian field, characterized by the power spectrum ${\cal P}_{\zeta}(k)$, defined as
\begin{equation}
    \langle \zeta(\mathbf{k}) \, \zeta(\mathbf{k'})   \rangle= \frac{2 \pi^2}{k^3} \mathcal{P}_{\zeta} (k) (2 \pi)^3 \delta^{(3)}(\mathbf{k}+\mathbf{k'}).
\end{equation} 
From the power spectrum one can define the various momenta
\begin{eqnarray}
\sigma_n^2&=&\int\frac{{\rm d}k}{k}\, k^{2n}\,{\cal P_\zeta}(k),
\end{eqnarray}
and in particular the quantity
\be
\langle k^4\rangle=\frac{\sigma_2^2}{\sigma_0^2},
\ee
which will be useful in the following.
The  average profile of the peak has been computed in Ref. \cite{Bardeen:1985tr}
\be
\label{profile}
{\zeta}_{\rm pk}({\bf x})=
\frac{\nu}{\sigma_0\left(1-\gamma^2\right)}\left(\xi+R_*^2\frac{\nabla^2 \xi}{3}\right)-\frac{x / \gamma}{\sigma_0\left(1-\gamma^2\right)}\left(\gamma^2 \xi+R_*^2\frac{\nabla^2 \xi}{3}\right)+R_*^2\frac{5}{2}\left(\frac{x}{\sigma_0\gamma}\right)\left(\frac{\xi^{\prime}}{r}-\frac{\nabla^2 \xi}{3}\right) A(e, p),
\ee
where 
\begin{equation}\label{Eq:psi}
    \xi(r)\equiv \langle \zeta({\bf x})\zeta(0)\rangle=\int \frac{{\rm d}k}{k}\,  \frac{\sin kr}{kr} \,{\cal P}_\zeta(k)
\end{equation}
is the curvature perturbation two-point correlator, 
$\nu=\zeta_0/\sigma_0$, $\zeta_0$ being the amplitude of the peak, $\gamma=\sigma_1^2/\sigma_2\sigma_0$ indicates the width of the power spectrum ($\gamma\simeq 0$ or 1 respectively for very peaked or flat spectra, respectively), $R_*=\sqrt{3}\sigma_1/\sigma_2$ measures the radius of the peaks and $x \equiv - \nabla^{2}  \zeta_{\rm pk}/\sigma_2$ (not to be confused with the position variable $x_k$). Finally 
\begin{equation}
\label{eq:A_function}
    A(e,p) = 3e \left[1- \sin^2 \theta (1+ \sin^2 \phi) \right]+p \left[1-3 \sin^2 \theta \cos^2 \phi \right] 
\end{equation}
measures the aspherical part of the peak in the convention of spherical coordinates. The parameters $e$ (``ellipticity") and $p$ (``prolateness") 
are in the range $0\leq e\leq1/2$ and $-e \leq p \leq e$.  
For instance, if 
$p>0$, 
the shape is oblate (pancake like) while it is prolate (cigar like) for $p<0$.  
It is also common to define ``spheroids" as ellipsoids with two equal eigenvalues. 

In the following it will prove convenient to  consider the
immediate neighbourhood  of $\zeta({\bf x})$ around a peak which we conveniently locate  at ${\bf x}= 0$. We  Taylor expand around it  up to  second order as
\be
\label{t}
\zeta_{\rm pk}({\bf x})=\zeta_0-\frac{1}{2}\sum_{p=1}^3\lambda_p\,x_p^2,
\ee
where the $\lambda_k$'s are the eigenvalues of the matrix 
$-\partial_i\partial_j\zeta$
and we have considered that the axes are oriented along
the  principal axes,   $x_k$ being the Cartesian coordinates. We can also arrange the $\lambda_k$'s so
that $\lambda_1\geq \lambda_2\geq \lambda_3$ by considering the rotation of the axes without loss of generality.
Around the peak, the contours of $\zeta_{\rm pk}({\bf x})$ are described by the ellipsoids 
given by $\zeta_0-\zeta_{\rm pk}({\bf x})\approx \sum_{k} x^2_k/(1/\lambda_k)={\rm constant.}$ with the radius along each axis being proportional to 
$1/\sqrt{\lambda_k}$. The shape of the ellipsoid is characterized by 
\begin{equation}
e = \frac{\lambda_1-\lambda_3}{2 \sum_i \lambda_i}, \, \,\,\,\,\,\, \,\, \,\, \,\, \,\,
p = \frac{\lambda_1-2 \lambda_2 +\lambda_3}{2 \sum_i \lambda_i},
\label{eq:elipsoid}
\end{equation}
which can be inverted to give 
\bea
    \lambda_1 &= \frac{x \sigma_2}{3}(1+3e +p),  \\ 
    \lambda_2 &= \frac{x \sigma_2}{3}(1-2p), \\ 
    \lambda_3 &= \frac{x \sigma_2}{3}(1-3e+p). 
    \label{eq:lambda_values_elipsoid}
\eea
The values of $e$ and $p$  
follow a specific probability distribution conditioned to
$x$ given by \cite{Bardeen:1985tr}
\bea
\label{eq:P_ep}
    P(e,p|x) &= \frac{3^{2} 5^{5/2}}{\sqrt{2\pi}} \frac{x^8}{f(x)}\exp\left[-\frac{5}{2} x^2 (3 e^2 + p^2)\right]W(e,p) , \\ 
    W(e,p) &= (1-2p)\left[ (1+p)^2 - (3 e )^2 \right] e (e^2-p^2) \chi(e,p),   
\eea
where $f(x)$ is given by
\be
\label{eq:f}
    f(x)=\frac{1}{2}x(x^2-3)\left(\erf\left[\frac{1}{2}\sqrt{\frac{5}{2}}x\right]+\erf \left[ \sqrt{\frac{5}{2}}x\right]\right) 
    +\sqrt{\frac{2}{5\pi}}\left(\mathcal{C}_1(x)\exp\left[-\frac{5}{8}x^2\right]+\mathcal{C}_2(x)\exp\left[-\frac{5}{2}x^2\right]\right)
\ee
with 
\begin{align}
    \mathcal{C}_1(x) = \frac{8}{5}+\frac{31}{4}x^2  \ ,  \, \, \, \,  \mathcal{C}_{2}(x) = -\frac{8}{5}+\frac{1}{2}x^2,  
\end{align}
and $\chi(e,p)$ is defined as
\begin{equation}
\left\{
\begin{aligned}
\chi &= 1 , \, \, \,  \, 0 \leq e \leq 1/4 , -e \leq p \leq e, \\
\chi &= 1 , \, \, \,  \, 1/4 \leq e \leq 1/2 , -(1-3e) \leq p \leq e, \\
\chi &= 0 , \, \, \,  \, \textrm{otherwise.}
\end{aligned}
\right.
\end{equation}
This expression does not explicitly depend on $\nu$, and
the domain with a non-vanishing value is restricted to $|p| \leq e$ and $e \geq 0$. Specifically, the allowed domain  $(e,p)$ is the interior of a triangle bounded by the points $(0,0)$, $(1/4,-1/4)$ and $(1/2,1/2)$.

Finally, in terms of the shape parameter $\gamma$, we can define the conditional probability of $x$ given $\nu$ as\,\cite{Bardeen:1985tr}
\be
P(x|\nu)=\frac{e^{-(x-\gamma\nu)^2/2(1-\gamma^2)}}{\sqrt{2\pi(1-\gamma^2)}}\frac{f(x)}{G(\gamma,\gamma\nu)},
\ee
where
\be
G(\gamma, \gamma \nu) = \int_0^\infty \dd x \,f(x) \,\frac{e^{-(x-\gamma\nu)^2/2(1-\gamma^2)}}{\sqrt{2\pi(1-\gamma^2)}}.
\ee
An important remark at this stage is the following. It is clear that, in making the statement that $\zeta({\bf x})$ has a peak  at some point, one 
implicitly assumes a finite resolution scale. Since the power spectrum of the curvature perturbation typically  increases as one
goes to smaller scales, a given maximum (or minimum) found at some
resolution scale would, most probably, disappear in making the
resolution finer. Thus, one should first define a resolution scale $\Rs$
and then the structure of the curvature field. For a given comoving
(spatial) resolution scale $\Rs$, one needs to introduce   a filtered curvature perturbation
defined by its truncated power spectrum

\be
{\cal P}_\zeta(k,\Rs)={\cal P}_\zeta(k) W^2(k,\Rs).
\ee
 From now on, we will intend the  curvature power spectrum to be the filtered one, even though for notational simplicity we will not write explicitly its dependence on the filter scale. As we shall see, the most natural choice for the comoving length $\Rs$ is of the order of the comoving Hubble radius when a given wavenumber re-enters the horizon.

\section{Understanding the role of peaks  in  the SIGWs}
\label{Sec::SIGW_Nature}
We are now in the position to assess how much peaks contribute to the SIGW background.  Consider a given peak of the Bardeen potential $\Psi$. In coordinate space it is given by 
\be
\Psi_{\rm pk}({\bf x},\eta)=\frac{2}{3}\int{\rm d}^3y \,T(y,\eta)\zeta_{\rm pk}({\bf x}-{\bf y}),
\ee
where $T(y,\eta)$ is the corresponding radiation transfer function. 
Since the GWs are sensitive to the quadrupole of the configuration, we first  focus  on the  aspherical part of the peak. To deal with it, it proves convenient to expand in Taylor around ${\bf x}-{\bf y}=0$ following Eq. (\ref{t}) 
\be
\label{Taylor}
\zeta_{\rm pk}({\bf x}-{\bf y})=\zeta_0-\frac{1}{2}\sum_{p=1}^3\lambda_p(x-y)_p^2.
\ee
so that we can easily extract the aspherical part  
\be
\Psi_{\rm pk}^{\rm asph}({\bf x},\eta)=-\frac{1}{2}\sum_{p=1}^3\lambda_p\int{\rm d}^3y \,\frac{2}{3}T(y,\eta)(x-y)_p^2.
\ee
Its Fourier transform reads
\begin{equation}
\begin{aligned}
\Psi_{\rm pk}^{\rm asph}({\bf k},\eta)&=-\frac{1}{2}\sum_{p=1}^3\lambda_p
\int{\rm d}^3x\, e^{-i{\bf k}\cdot {\bf x} }
\int{\rm d}^3y \,\frac{2}{3}T(y,\eta)(x-y)_p^2\\
&=-\frac{1}{2}\sum_{p=1}^3\lambda_p
\int{\rm d}^3x\, e^{-i{\bf k}\cdot {\bf x} }
\int{\rm d}^3y\int\frac{{\rm d}^3 k'}{(2\pi)^3}e^{i{\bf k}'\cdot {\bf y}} \frac{2}{3}T(k',\eta)(x-y)_p^2\\
&=-\frac{1}{2}\sum_{p=1}^3\lambda_p
\int{\rm d}^3x\, e^{-i{\bf k}\cdot {\bf x} }
\int{\rm d}^3z\int\frac{{\rm d}^3 k'}{(2\pi)^3}e^{i{\bf k}'\cdot ({\bf x}-{\bf z})} \frac{2}{3}T(k',\eta)z_p^2\\
&=-\frac{1}{2}\sum_{p=1}^3\lambda_p
\int{\rm d}^3z\, e^{-i{\bf k}\cdot {\bf z}} \frac{2}{3}T(k,\eta)z_p^2\\
&=\frac{1}{2}\sum_{p=1}^3\lambda_p\,\frac{2}{3}T(k,\eta) \nabla^2_k \int{\rm d}^3z\, e^{-i{\bf k}\cdot {\bf z}}\\
&=\frac{1}{2}\sum_{p=1}^3\lambda_p\,\frac{2}{3}T(k,\eta)(2\pi)^3 \frac{\partial^2}{\partial k_{\rm p}^2}\delta(\mathbf{k}).
\end{aligned}
\end{equation}
Consider now  the first term 
in the source (\ref{eq: source x space}). 
It provides the following expression (counting the symmetry factor)
\begin{align}
{\cal S}^s(\eta, \vk)&\supset
2\cdot 4\int\frac{{\rm d}^3 k'}{(2\pi)^3}e^{s,ij}({\bf k})\,k'_i k'_j\, 2\Psi (|{\bf k}-{\bf k}'|,\eta)\Psi_{\rm pk}^{\rm asph} ({\bf k}',\eta)  \nonumber\\
&=2\cdot 4\cdot\frac{4}{9}\int\frac{{\rm d}^3 k'}{(2\pi)^3}e^{s,ij}({\bf k})\,k'_i k'_j\, 2T(|{\bf k}-{\bf k}'|,\eta)\zeta (|{\bf k}-{\bf k}'|)\frac{1}{2}\sum_{p=1}^3\lambda_p\,T(k',\eta)(2\pi)^3  \frac{\partial^2}{\partial k_{\rm p}^2}\delta({\bf k}').
\end{align}
Choosing conveniently the vector ${\bf k}$ along the third axis, we have
\be
e^{+}_{ij}=
\left(
\begin{array}{cc}
1 & 0\\
0 & -1
\end{array}\right)_{ij}\,\,\,\,{\rm and} \,\,\,\, e^{\times}_{ij}=
\left(
\begin{array}{cc}
0 & 1\\
1 & 0
\end{array}\right)_{ij},
\ee
so that, integrating by parts,  
\begin{equation}
2\cdot 4\cdot\frac{4}{9}
\int\frac{{\rm d}^3 k'}{(2\pi)^3}(\lambda_1-\lambda_2)2T(|{\bf k}-{\bf k}'|,\eta)\zeta ({\bf k}-{\bf k}')T(k',\eta)\delta({\bf k}')(2\pi)^3=
2\cdot4\cdot\frac{4}{9}
(\lambda_1-\lambda_2)2T(k,\eta)\zeta({\bf k}).
\end{equation}
Repeating the same calculation for the whole source (\ref{eq: source x space}) for $w=1/3$, we get 
\begin{equation}
h_{\vk}(\eta)=\left[\frac{1}{a(\eta)} \int^\eta \dd \etat \,
  g_{\vk}(\eta, \etat) \, a(\etat)f(k,0,\eta) \right]\frac{8}{9}(\lambda_1-\lambda_2)\zeta ({\bf k})\equiv G(k,\eta)\frac{8}{9}(\lambda_1-\lambda_2)\zeta({\bf k}),
\end{equation}
where  \cite{Espinosa:2018eve}
\be
f(k_1,k_2,\eta)=4\left[2T(k_1\eta)T(k_2\eta)+\left(T(k_1\eta)+\frac{T'(k_1\eta)}{{\cal H}}\right)\left(T(k_2\eta)+\frac{T'(k_2\eta)}{{\cal H}}\right)\right],
\ee
and 
\be
f(k,0,\eta)=12\left[T(k,\eta)+\frac{1}{3}\frac{T'(k,\eta)}{{\cal H}}\right]=12\sqrt{3} \frac{\textrm{sin}(k\eta/\sqrt{3})}{k\eta}.
\ee
The comoving curvature perturbation  $\zeta({\bf k})$ around the peak has two contributions: the zero mode $\zeta_0$, which leads to another delta function $\delta({\bf k})$ and therefore we neglect it, and the fluctuation around it, which we keep. 
We can then easily calculate the GW power spectrum to be 
\be
\mathcal{P}_h(k,\eta)=
\frac{64}{81} \langle {\cal A}(e,p)\rangle\overline G^2(k,\eta){\cal P}_\zeta(k,\Rs),
\ee
where 
\be
\overline G^2(k,\eta)=\frac{648}{k^6\eta^2}
\ee
has been obtained by  averaging over the fast oscillations with frequency $k\eta\gg 1$, 
and 
\be
 {\cal A}(e,p)=\left[(\lambda_1-\lambda_2)^2+(\lambda_2-\lambda_3)^2+(\lambda_3-\lambda_1)^2\right]=\frac{2}{3}x^2\sigma_2^2(3 e^2+p^2),
\ee
which we have 
averaged   over the probability of the aspherical parameters $e$ and $p$, with $\langle p^2\rangle\ll \langle e^2\rangle$. We finally obtain the amount of SIGW at the time of emission
\begin{equation} \label{eq:fund)}  
    \OGW(k)|_{\rm e} \simeq 
\frac{128}{3}
    \langle e^2 x^2\rangle
    \frac{\langle k^4\rangle}{k^4}\sigma_0^2{\cal P}_\zeta(k)  
   W^2(k,\Rs).
\end{equation}
This is our main result. 
The expression for the GW spectrum, as expected, is proportional to the square of the smoothed scalar power spectrum and to the average eccentricity, showing explicitly how the asphericity of the peaks is the GW source.

In our analysis below  we will  adopt a Gaussian window function,  i.e. $W(k,\Rs)\equiv{\rm exp}(-\Rs^2k^2/2)$ to define the smoothing scale. 
This choice is motivated by the well--known drawbacks of both the real--space top--hat 
and the sharp $k$-space filters\,\cite{Bardeen:1985tr}\footnote{The sharp cutoff in either real or Fourier space 
produces spurious oscillations in the correlation functions, which extend far beyond the  cutoff scale and contaminate the physical signal. By contrast, the Gaussian filter provides  an optimal balance between smoothing and localization: it suppresses small--scale fluctuations  without introducing artificial oscillations, and it allows for a clean identification of the  relevant peaks in the curvature perturbation. Physically, this corresponds to the idea that  the collapse of overdensities is not determined by a sharp boundary in real space, but rather  by a gradual weighting of modes around the horizon scale.}. Furthermore, we take the  smoothing radius to be of the order of the comoving horizon scale at which a given wavenumber enters the Hubble radius, $\Rs = R_H(k) \simeq  \eta_k = 1/k$.
\begin{figure}[h!]\centering
\includegraphics[width=0.7\columnwidth]{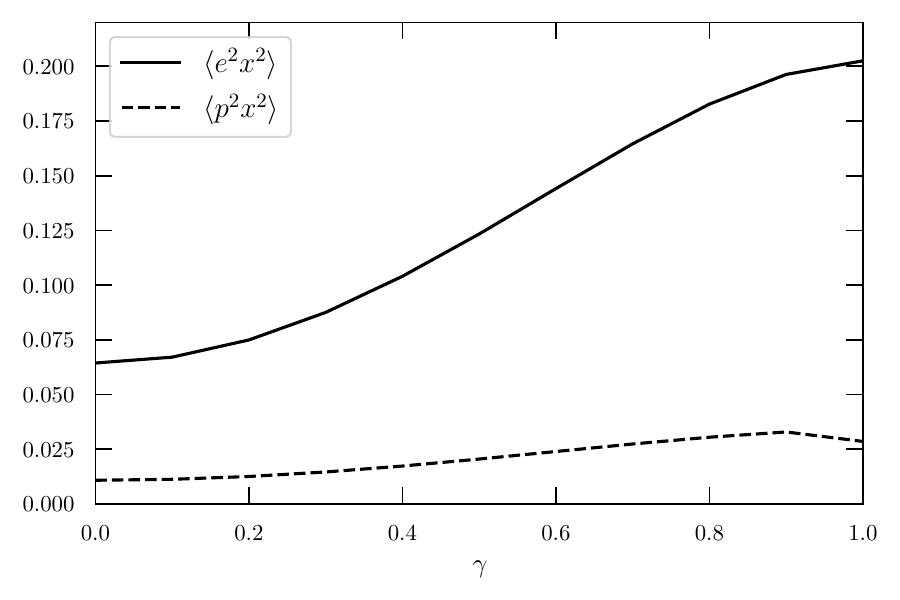}\vspace*{-1em}
\caption{Plot of $\langle e^2 x^2\rangle$ (solid) and $\langle p^2 x^2\rangle$ (dashed) as a function of $\gamma$.}
\label{fig:Finestra}
\end{figure}
\newline
We show in Fig.~\ref{fig:Finestra}  $\langle e^2 x^2\rangle$ and $\langle p^2 x^2\rangle$ as function of $\gamma$ where only peaks with $|\nu|\gsim 1$ have been considered \cite{Hoffman:1985pu}. 
%
Let us evaluate our result in a couple of  relevant cases. 
\vskip 0.2cm
\centerline{\it Peaked power spectrum}
\vskip 0.2cm
\noindent
For a peaked spectrum, we consider a Dirac delta 
$\Pz(k)=A_{\rm p} k_*\delta(k-k_*)$, 
for which
\be
\langle k^4\rangle=  k_*^4\,\,\,{\rm and}\,\,\, \langle e^2 x^2\rangle\simeq 0.2.
\ee
If we follow Ref. \cite{Pi:2020otn} and smooth $\OGW(k_{\rm p})|_{\rm e}(k)$ around $k_*$, we find
\be
\int\frac{{\rm d}k}{k}\OGW(k)|_{\rm e}=\frac{128}{3}\langle e^2 x^2\rangle e^{-2k_*^2 R^2_s}A^2_{\rm p}.
\ee
Taking $R_s$ to the comoving Hubble scale when the wavenumber $k_*$ enters the horizon, 
we find that the contribution to the GW background sourced by the peaks is  equal to 
$\OGW \approx 1.1\, A_{\rm p}^2$ in good agreement with the value reported  in Ref. \cite{Pi:2020otn} for $k=k_*$. Doing the same average with a lognormal power spectrum 
$(A_{\rm p}/\sqrt{2\pi\Delta^2}){\rm exp}\left[-{\rm ln}^2(k/k_*)/2\Delta^2\right]$
around the peak of the numerically GW spectrum $k_{\rm p}=2k_*/\sqrt{3}$ reported in Ref. \cite{Kohri:2018awv} we find
\be
\label{eq:OGW peaked}
\frac{1}{2\Delta}\int_{k_{\rm p} e^{-\Delta}}^{k_{\rm p} e^{\Delta}}\frac{{\rm d}k}{k}\OGW(k)|_{\rm e}\approx 1.6\, A_{\rm p}^2,
\ee
for $R_s=1/k_{\rm p}$  and  $\Delta=10^{-1}$, to be compared to the value $\OGW \approx 2.6\, A_{\rm p}^2$ reported in Ref. \cite{Pi:2020otn}.
\vskip 0.2cm
\centerline{\it Flat power spectrum}
\vskip 0.2cm
\noindent
In the case of a flat power spectrum 
defined by a band between $k_m$ and  $k_M\gg k_m$ and 
with a constant  amplitude $A_{\rm b}$, we obtain
\be
\langle k^4\rangle\sigma_0^2 = \frac{A_{\rm b}}{2 R_s^4}.
\ee
Now,  $\langle e^2  x^2\rangle \simeq 0.06$  for $\gamma\simeq 0$, and we choose    $\Rs \simeq \eta_k=1/k$ to be the comoving Hubble radius when the wavenumber $k$ enters the horizon, to obtain the estimate
\be
\label{eq:OGW flat}
\OGW(k)|_{\rm e}  \simeq  \frac{128}{3} \cdot \frac{e^{-1}}{2}\cdot 0.06\,A_{\rm b}^2\simeq 0.5 \,A_{\rm b}^2,
\ee
not far from the result quoted in  Refs.~\cite{Kohri:2018awv,Pi:2020otn}, $\OGW(k)|_{\rm e}  \simeq 0.8 \, A_{\rm b}^2$,  for a scale-invariant power spectrum.

In both the examples we have provided, the contribution from the peaks is about 60\% of the full numerical standard result.
Indeed we notice that Eqs.~\eqref{eq:OGW peaked} and \eqref{eq:OGW flat} can be reproduced by multiplying it by 0.6: $2.6\cdot 0.6\simeq $ 1.6 and $0.8\cdot 0.6\simeq 0.5$ for peaked and flat power spectra, respectively. 
Let us see now where the 60\% comes from.
It is well-known that peaks are more correlated that the underlying fluctuating field. Approximating  the correlation of peaks in terms of the correlation of the curvature perturbation as \cite{Kaiser:1984sw}
\be
\xi_{\rm pk}({\bf x})\simeq \nu^2\,\xi({\bf x}),
\ee
we immediately find 
\be
{\cal P}_{\rm pk}(k)\simeq \nu^2\,{\cal P}_\zeta(k),
\ee
and therefore to estimate the contribution of peaks one can take the standard second-order calculation for the SIGW, use ${\cal P}_{\rm pk}(k)$ instead of ${\cal P}_\zeta(k)$ and weight it by the factor 
\be
\label{fraction}
\left(2\int_1^\infty\frac{{\rm d}\nu}{\sqrt{2\pi}}\nu^2e^{-\nu^2/2}\right)^2\simeq 0.6.
\ee
The 2 comes from counting all peaks with $|\nu|\gsim 1$ and the square from the fact one should count the probability that each of the two peaks in the correlation is above the threshold. This estimate further supports the idea that SIGWs are generated mainly around peaks.

\section{Conclusions}
\label{Sec::Conclusions}
The detection of a SIGW background, e.g. through the forthcoming LISA experiment \cite{LISACosmologyWorkingGroup:2025vdz}, will open a new window towards understanding the evolution of the universe at its infancy. In this paper we have provided an  interpretation of the origin of the SIGW
background by estimating the amount of GWs generated   around the peaks of the scalar
perturbations. Our calculation is definitely not conclusive. One natural avenue to pursue is to abandon the simplifying hypothesis of gaussianity. Indeed, SIGWs may be also  a powerful tools to investigate non-Gaussianity in the early universe since their production is  sensitive to the statistical properties of scalar curvature fluctuations. Fully understanding peaks in a non-Gaussian theory is however not a done task, and therefore we have not attempted to extend our calculations to the non-Gaussian case. We intend to do so in future work.

\vskip 0.8cm
\centerline{\bf Acknowledgments}
\vskip 0.2cm
\noindent
We thank V. De Luca for useful discussions and comments on the draft.
G.P. acknowledges support by ASI Grant No. 2016-24-H.0 and by Fondazione Angelo Della Riccia and Fondazione Aldo Gini.
D.R.~is supported at U.~of Zurich by the UZH Postdoc Grant 2023 Nr.\,FK-23-130.
A.R. acknowledges support from the Swiss National Science Foundation (project number CRSII5\_213497) and from  the Boninchi Foundation for the project ``PBHs in the Era of GW Astronomy''. 
\printbibliography

\end{document}